\documentclass[conference,onecolumn,letterpaper]{IEEEtran}
\IEEEoverridecommandlockouts
\usepackage[compact]{titlesec}
\titlespacing{\section}{0pt}{2ex}{1ex}
\titlespacing{\subsection}{0pt}{1ex}{0ex}
\titlespacing{\subsubsection}{0pt}{0.5ex}{0ex}
    
\usepackage{paralist}

\usepackage{cite}
\usepackage{tabularx}
\usepackage[left=1.72cm,right=1.72cm,top=0.71in,bottom=0.96in]{geometry}

\usepackage[cmex10]{amsmath} 

\usepackage{amssymb,amsfonts,mathtools,bbm}

\usepackage{physics}

\usepackage{algorithmic}
\usepackage{graphicx}
\usepackage{textcomp}
\usepackage{xcolor}
\usepackage{adjustbox}
\usepackage{graphicx}
\usepackage{booktabs} 

\usepackage{float}
\usepackage{adjustbox}

\usepackage{tikz}
\tikzset{>=latex}

\usetikzlibrary{
quantikz,
shapes, 
arrows,
positioning,
intersections, 3d,perspective 
}

\usepackage{pgf}
\usepackage{pgfplots}
\pgfplotsset{compat=1.3}   
\usepgfplotslibrary{colorbrewer}

\usepgfplotslibrary{external}

\PassOptionsToPackage{hyphens}{url}
\usepackage[bookmarks=false]{hyperref}
\usepackage{cite}
\usepackage[capitalize]{cleveref}

\usepackage{xcolor}
\usepackage[
shortcuts,       
acronym,         
nomain,          
section=chapter, 
toc=true,        
]{glossaries}
\glsenableentrycount 

\definecolor{mypurple}{rgb}{0.49,0.18,0.56}
\definecolor{mygold}{rgb}{0.93,0.49,0.13}
\definecolor{mygreen}{rgb}{0,0.5,0}
\definecolor{myblue}{rgb}{0,0,0.75}
\definecolor{mymagenta}{cmyk}{0,1,0,0.12}
\definecolor{mygray}{rgb}{0.5,0.5,0.5}

\usepackage{braket}
\usepackage{adjustbox}

\usepackage{changepage}

\usetikzlibrary{
quantikz, 
shapes,  
arrows,  
}

\DeclareUnicodeCharacter{0391}{\Alpha}       
\DeclareUnicodeCharacter{0392}{\gamma}        
\DeclareUnicodeCharacter{0393}{\Gamma}       
\DeclareUnicodeCharacter{0394}{\Delta}       
\DeclareUnicodeCharacter{0395}{\Epsilon}     
\DeclareUnicodeCharacter{0396}{\Zeta}        
\DeclareUnicodeCharacter{0397}{\Eta}         
\DeclareUnicodeCharacter{0398}{\Theta}       
\DeclareUnicodeCharacter{0399}{\Iota}        
\DeclareUnicodeCharacter{039A}{\Kappa}       
\DeclareUnicodeCharacter{039B}{\Lambda}      
\DeclareUnicodeCharacter{039C}{\Mu}          
\DeclareUnicodeCharacter{039D}{\Nu}          
\DeclareUnicodeCharacter{039E}{\Xi}          
\DeclareUnicodeCharacter{039F}{\Omicron}     
\DeclareUnicodeCharacter{03A0}{\Pi}          
\DeclareUnicodeCharacter{03A1}{\Rho}         
\DeclareUnicodeCharacter{03A3}{\Sigma}       
\DeclareUnicodeCharacter{03A4}{\Tau}         
\DeclareUnicodeCharacter{03A5}{\Upsilon}     
\DeclareUnicodeCharacter{03A6}{\Phi}         
\DeclareUnicodeCharacter{03A7}{\Chi}         
\DeclareUnicodeCharacter{03A8}{\Psi}         
\DeclareUnicodeCharacter{03A9}{\Omega}       
\DeclareUnicodeCharacter{03B1}{\alpha}       
\DeclareUnicodeCharacter{03B2}{\beta}        
\DeclareUnicodeCharacter{03B3}{\gamma}       
\DeclareUnicodeCharacter{03B4}{\delta}       
\DeclareUnicodeCharacter{03B5}{\epsilon}     
\DeclareUnicodeCharacter{03B6}{\zeta}        
\DeclareUnicodeCharacter{03B7}{\eta}         
\DeclareUnicodeCharacter{03B8}{\theta}       
\DeclareUnicodeCharacter{03B9}{\iota}        
\DeclareUnicodeCharacter{03BA}{\kappa}       
\DeclareUnicodeCharacter{03BB}{\lambda}      
\DeclareUnicodeCharacter{03BC}{\mu}          
\DeclareUnicodeCharacter{03BD}{\nu}          
\DeclareUnicodeCharacter{03BE}{\xi}          
\DeclareUnicodeCharacter{03BF}{\omicron}     
\DeclareUnicodeCharacter{03C0}{\pi}          
\DeclareUnicodeCharacter{03C1}{\rho}         
\DeclareUnicodeCharacter{03C3}{\sigma}       
\DeclareUnicodeCharacter{03C2}{\varsigma}    
\DeclareUnicodeCharacter{03C4}{\tau}         
\DeclareUnicodeCharacter{03C5}{\upsilon}     
\DeclareUnicodeCharacter{03D5}{\phi}         
\DeclareUnicodeCharacter{03C6}{\varphi}      
\DeclareUnicodeCharacter{03C7}{\chi}         
\DeclareUnicodeCharacter{03C8}{\psi}         
\DeclareUnicodeCharacter{03C9}{\omega}       

\newtheorem{Remark}{Remark}

\newtheorem{Theorem}{Theorem}

\newtheorem{lemma}{Lemma}

 \newtheorem{Lemma}[lemma]{Lemma}

\usepackage[utf8]{inputenc} 
\usepackage[T1]{fontenc}
\usepackage{url}
\usepackage{ifthen}
\usepackage{cite}

\newtheorem{Definition}{Definition}

\newtheorem{Corollary}{Corollary}
\newtheorem{Notation}{Notation}

\begin{document}

\title{Galaxy Codes: Advancing Achievability for Deterministic Identification via Gaussian Channels}
%
%
\author{%
\IEEEauthorblockN{%
Holger Boche\IEEEauthorrefmark{1}
Christian Deppe\IEEEauthorrefmark{2}
Safieh Mahmoodi\IEEEauthorrefmark{3}
Golamreza Omidi\IEEEauthorrefmark{2}\IEEEauthorrefmark{3}
}
\IEEEauthorblockA{%
\IEEEauthorrefmark{1}
Technical University of Munich,
Chair of Theoretical Information Technology,
80290 Munich, Germany\\
\IEEEauthorrefmark{2}
Institute for Communications Technology, Technische Universität Braunschweig\\
\IEEEauthorrefmark{3}
Isfahan University of Technology, Department of Mathematical
Sciences, Isfahan, 84105, Iran\\
Email:
boche@tum.de, christian.deppe@tu-bs.de, mahmoodi@iut.ac.ir,  romidi@iut.ac.ir }
}

\maketitle

\begin{abstract}
Deterministic identification offers an efficient solution for scenarios where decoding entire messages is unnecessary. It is commonly used in alarm systems and control systems. A key advantage of this approach is that the capacity for deterministic identification in Gaussian channels with power constraints grows superexponentially, unlike Shannon’s transmission capacity. This allows for a significantly higher number of messages to be transmitted using this event-driven method.
So far, only upper and lower bounds for deterministic identification capacity have been established. Our work introduces a novel construction: galaxy codes for deterministic identification. Using these codes, we demonstrate an improvement in the achievability bound of $\frac 14$ to $\frac 38$, representing a previously unknown advance that opens new possibilities for efficient communication.
\end{abstract}


\section{ Introduction}

In modern times, both the number of messages exchanged and the number of communication partners in networks are steadily increasing. At the same time, there is a continuous effort to enhance communication performance by expanding capacities. However, such approaches face technical limitations. A promising alternative is goal-oriented communication, as described in the Post-Shannon Theory.

One specific implementation of this approach is message identification introduced in \cite{Ahlswede1989}. Unlike traditional decoding in the sense of Shannon \cite{Shannon1948}, message identification focuses on the receiver determining whether a specific, sender-unknown message is present. This type of modulation is particularly useful in applications like alarm or control systems. Practical use cases include the concept of the digital twin or molecular communication.

Deterministic identification was initially introduced in the context of communication complexity in \cite{jaja1985identification}. Later, the concept of randomized identification was explored for discrete memoryless channels (DMCs) in \cite{Ahlswede1989}, where messages are processed using randomized encoding. It was shown that for DMCs, the number of identifiable messages grows double-exponentially with the block length, in contrast to traditional transmission, where growth is merely exponential.

In practical scenarios, however, randomized encoding is not always feasible. An alternative is a deterministic identification, which also allows for an increase in the number of messages, though the growth, in this case, is only exponential with respect to the block length \cite{salariseddigh2021deterministic}.

For applications in wireless communication, the Gaussian channel becomes particularly relevant. It has been demonstrated that, for an additive white Gaussian noise (AWGN) channel with power constraints \cite{Salariseddigh2020}, deterministic identification grows superexponentially with block length. However, thus far, only bounds for the deterministic identification capacity have been derived.

To enhance the best-known achievability bound for deterministic identification, we propose Galaxy codes. The design of Galaxy codes is inspired by a fractal-like hierarchy found in celestial systems: moons orbiting planets, planets orbiting stars, stars orbiting the center of a galaxy, and galaxies orbiting the center of a cluster; hence the name. At each level of this hierarchy, the structure is represented by points on a hypersphere centered around a specific point, which then serves as the origin for hyperspheres at the next level.

The decision-making process mirrors this hierarchical structure, proceeding from the top level down. Starting with a given codeword, the process first determines whether it belongs to its highest-level galaxy (“yes” decision) or not (“no” decision). A “yes” advances the process to decide its membership in a specific star system, and then, if confirmed, to its planet. At the final level, the decision focuses on distinguishing the codeword itself (a “moon”) from other moons orbiting the same planet.

This hierarchical approach effectively zooms in through successive scales to pinpoint a single codeword. To successfully identify a message, the binary test at each level must result in a “yes.” A single “no” at any stage halts the process, rejecting the codeword.

This paper is structured as follows: Section~\ref{preliminaries} introduces the notations, definitions, and the system model used throughout the paper. Section~\ref{results} surveys the known results for message identification via the Gaussian channel and presents our new achievability bound. Section~\ref{galaxy} focuses on the construction of a novel identification code, referred to as the Galaxy code. Section~\ref{properties} analyzes the properties of Galaxy codes, including code size, code distance, and error performance. Section~\ref{proof} provides the proof of our main result. Finally, we conclude in Section~\ref{conclusions}.

\section{Preliminaries and Notations}\label{preliminaries}

We use bold symbols, $\mathbf{u}$, to denote points $\mathbf{u} = (u_1, u_2, \ldots, u_n)$ in $\mathbb{R}^n$. 
Additionally, bold symbols with arrows, $\vec{\mathbf{u}}$, represent vectors $\vec{\mathbf{ou}} = (u_1, u_2, \ldots, u_n)$ in $\mathbb{R}^n$, 
where $\mathbf{o} = (0, 0, \ldots, 0)$ is the origin.

The uppercase symbols $X$ and $Y$, denote random variables, while the lowercase symbols, $x$ and $y$, represent constant real numbers. 
The notation $\log$ refers to the logarithm to base 2, also known as the binary logarithm.

Let $\|\mathbf{u}\|_2$ denote the Euclidean 2-norm, defined as 
\[
\|\mathbf{u}\|_2 = \left[\sum_{i=1}^n u_i^2\right]^{1/2}.
\]

By $S(\mathbf{u}, r)$, we refer to an $n$-dimensional sphere with center $\mathbf{u} \in \mathbb{R}^n$ and radius $r$.

Our system model considers a point-to-point communication channel consisting of a sender, a noisy channel, and a receiver. 
The sender transmits codewords of block length \( n \) to the receiver, who observes \( n \) noisy symbols.

Let \( \mathcal{M} := \{1, 2, \ldots, N\} \) represent a set of messages. Define \( \mathcal{X} \) and \( \mathcal{Y} \) as the input and output sets of the channel, respectively. 
Each message \( i \in \mathcal{M} \) is encoded into a codeword \( \mathbf{u}_i = (u_{i1}, u_{i2}, \ldots, u_{in}) \in \mathcal{X}^n \). 
After passing through the channel, the output is represented by a vector \( \mathbf{y} = (y_1, y_2, \ldots, y_n) \in \mathcal{Y}^n \).

In the identification problem, for each message \( i \in \mathcal{M} \), the decoder must determine whether or not message \( i \) was sent based on the received output. 
A collection \( \{D_i: i \in \mathcal{M}\} \) is referred to as the set of decoding sets. 
If the channel output lies in \( D_i \), the decoder decides that \( i \) was sent; otherwise, it concludes that \( i \) was not sent.

In this paper, we consider a Gaussian channel \( \mathcal{G} \) characterized by the input-output relationship:
\[
\mathbf{Y} = \mathbf{u}_i + \sigma \mathbf{Z},
\]
where \( \mathbf{Y} = (Y_1, Y_2, \ldots, Y_n) \) is the channel output, \( \mathbf{u}_i = (u_{i1}, u_{i2}, \ldots, u_{in}) \) is the channel input, and \( \mathbf{Z} = (Z_1, Z_2, \ldots, Z_n) \) is a noise vector. 
The noise components \( Z_1, Z_2, \ldots, Z_n \) are independent and identically distributed (i.i.d.) standard normal random variables, \( Z_i \sim \mathcal{N}(0, 1) \).

The standard normal probability density function \( \phi(z) \) and cumulative distribution function \( \Phi(x) \) are defined as:
\[
\phi(z) = \frac{1}{\sqrt{2\pi}} e^{-z^2 / 2}, \quad \forall z \in \mathbb{R},
\]
\[
\Phi(x) = \int_{-\infty}^x \phi(z) \, dz, \quad \forall x \in \mathbb{R}.
\]

For a set \( D \subseteq \mathbb{R}^n \) and \( \mathbf{u} \in \mathbb{R}^n \), the probability that the output of the Gaussian channel \( \mathcal{G} \) lies in \( D \) when the input is \( \mathbf{u} \) is denoted by \( P(D \mid \mathbf{u}) \).

\begin{Definition}
A deterministic $(n,N=2^{nR\log n},\lambda _{1},\lambda _{2})$ DI
(deterministic identification code) for Gaussian channel $\mathcal{G}$, is
defined as a pair $(\mathcal{U},\mathcal{D})$\bigskip , where $\mathcal{U}=\{%
\mathbf{u}_{i},i\in \mathcal{M}\}\subseteq $ $\mathcal{X}^{n}$ with power
constraint $\left\Vert \mathbf{u}_{i}\right\Vert
_{2}^{2}=\sum\limits_{k=1}^{n}u_{ik}^{2}\leq nP$ and collection of decoding
sets $\mathcal{D}=\left\{ D_{i}\subset \mathbb{R}^{n},\text{ }i\in \mathcal{M%
}\right\} $. The error probabilities of types I and II of this system equal
to 
\begin{align*}
& P(\text{output}\notin D_{i}|\text{sending a codeword corresponding to }%
i)\\
= & P(D_{i}^{c}|\mathbf{u}_{i})\leq \lambda _{1}~~~~~\forall i\in \mathcal{M}
\end{align*}%
\begin{align*}
& P(\text{output}\in D_{j}|\text{sending a codeword corresponding to }%
i)\\
= & P(D_{j}|\mathbf{u}_{i})\leq \lambda _{2}~~~~~~~~\forall i\neq j\in 
\mathcal{M}
\end{align*}
\end{Definition}

\begin{Definition}
The rate $R$ is called achievable if for every $\lambda _{1}$ and $\lambda
_{2}>0$ and large enough $n$, there exists a $(n,N=2^{nR\log n},\lambda
_{1},\lambda _{2})$ DI code over the Gaussian channel $\mathcal{G}$ and the
capacity of the channel equals the supremum of all achievable rates and will
be denoted by $C_{DI}(\mathcal{G})$.\bigskip
\end{Definition}

\section{Known results and main result}\label{results}

The paper \cite{salariseddigh2021deterministic} was the first to provide a detailed analysis of deterministic identification over Gaussian channels with power constraints. 
Before this, it was not known that in this case, the scaling of messages for deterministic identification differs from that of message transmission over these channels. 
This result was highly unexpected. Specifically, the following was shown, where $L(n,R)$ denote the scaling:

\begin{Theorem}[\cite{salariseddigh2021deterministic}]
The deterministic identification (DI) capacity of the Gaussian channel \( \mathcal{G} \) with power constraints in the \( 2^{n \log n} \)-scale, i.e., for \( L(n, R) = 2^{(n \log n) R} \), is bounded by:
\[
\frac{1}{4} \leq C_{\text{DI}}(G, L) \leq 1. \tag{92}
\]
As a result, the DI capacity is infinite in the exponential scale and zero in the double exponential scale. Formally,
\[
C_{\text{DI}}(G, L) =
\begin{cases} 
\infty, & \text{for } L(n, R) = 2^{nR}, \\
0, & \text{for } L(n, R) = 2^{2^{nR}}.
\end{cases}
\]
\end{Theorem}

The work \cite{vorobyev2024deterministicidentificationcodesfading} improved the upper bound and established the following result:

\begin{Theorem}[\cite{vorobyev2024deterministicidentificationcodesfading}]
For a Gaussian AWGN channel \( \mathcal{G} \) with power constraints for \( L(n, R) = 2^{(n \log n) R} \), the deterministic identification (DI) capacity satisfies:
\[
C_{\text{DI}}(\mathcal{G}) \leq \frac{1}{2}.
\]
\end{Theorem}

In this work, we focus on improving the lower bound. 
To achieve this, we developed a new coding method called Galaxy Codes. 
It turns out that this method allows us to establish a better lower bound. Specifically, we prove the following result:

\begin{Theorem}\label{mainresult}
For a Gaussian AWGN channel with power constraints, the deterministic identification (DI) capacity satisfies:
\[
C_{\text{DI}}(\mathcal{G}) \geq \frac{3}{8}.
\]
\end{Theorem}
The proof is given in Section~\ref{proof},

\section{Galaxy codes from spherical codes}\label{galaxy}

Before introducing the new ID code, we need some further notations.

\begin{figure}
    \centering
    \includegraphics[width=0.9\linewidth]{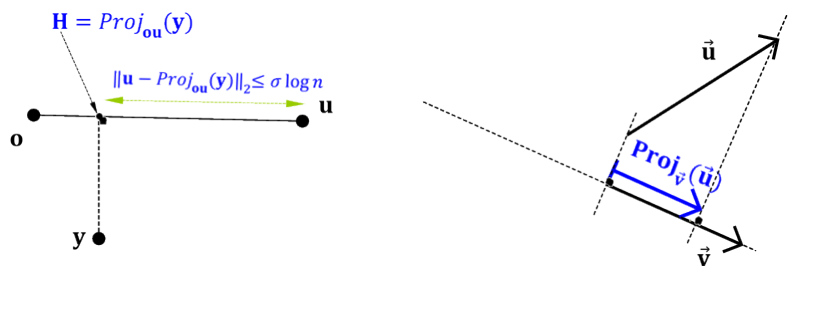}
    \caption{{\protect\footnotesize Projection of point and vector.}}\label{projection}
\end{figure}

\begin{Notation}
\label{Note1} Consider two arbitrary points $\mathbf{o}$ and $\mathbf{u}$ in 
$\mathbb{R}^{n}$. For each point $\mathbf{y}\in \mathbb{R}^{n}$ the
orthogonal projection of $\mathbf{y}$ onto the line $\mathbf{o}\mathbf{u}$
is the point that reaches the line $\mathbf{o}\mathbf{u}$ if we drop a
perpendicular from $\mathbf{y}$ to the line $\mathbf{o}\mathbf{u}$ and is
denoted by $Proj_{\mathbf{o}\mathbf{u}}(\mathbf{y})$. Let  
\begin{equation*}
P_{\mathbf{o},\mathbf{u}}=\{\mathbf{y}\in \mathbb{R}^n: \|\mathbf{u}-Proj_{%
\mathbf{o}\mathbf{u}}(\mathbf{y})\|_{2}\leq \sigma \log n \}.
\end{equation*}%
Consider two arbitrary vectors $\vec{\mathbf{u}}=(u_{1},u_{2},\ldots,u_{n})
$ and $\vec{\mathbf{v}}=(v_{1},v_{2},\ldots,v_{n})$ in $\mathbb{R}^{n}$. The
inner product of two vectors $\vec{\mathbf{v}}$ and $\vec{\mathbf{u}}$ in $%
\mathbb{R}^{n}$, is defined by $<\vec{\mathbf{u}},\vec{\mathbf{v}}%
>=\sum\limits_{i=1}^{n}u_{i}v_{i}$. The orthogonal projection of $\vec{%
\mathbf{u}}$ onto the vector $\vec{\mathbf{v}}$ is the vector $\frac{<\vec{%
\mathbf{u}},\vec{\mathbf{v}}>}{<\vec{\mathbf{v}},\vec{\mathbf{v}}>}\vec{%
\mathbf{v}}$ and is denoted by $Proj_{\vec{\mathbf{v}}}(\vec{\mathbf{u}})$. For an illustration, see Fig~\ref{projection}

For points $\mathbf{o}$, $\mathbf{u}$, $\mathbf{y_1}$ and $\mathbf{y_2}$ in $%
\mathbb{R}^{n}$, by an easy task one can easily find the coordinate of the
orthogonal projection of $\mathbf{y_1}+\mathbf{y_2}$ onto the line $\mathbf{o%
}\mathbf{u}$ in $\mathbb{R}^{n}$ as follows: 
\begin{equation*}
Proj_{\mathbf{o}\mathbf{u}}(\mathbf{y_1}+\mathbf{y_2})= Proj_{\mathbf{o}%
\mathbf{u}}(\mathbf{y_1})+ Proj_{\vec{\mathbf{o}\mathbf{u}}}(\vec{\mathbf{y_2}}),
\end{equation*}
where $+$ in the right-hand side of this equation is the summation between two $n
$-tuples in $\mathbb{R}^{n}$ and $Proj_{\mathbf{o}\mathbf{u}}(\mathbf{y_1})$
is the projection of the point $\mathbf{y_1}$ on the line $\mathbf{o}\mathbf{%
u}$ and $Proj_{\mathbf{o}\mathbf{u}}(\vec{\mathbf{y_2}})$ is the projection
of the vector $\vec{\mathbf{y_2}}$ on the vector $\vec{\mathbf{o}}\mathbf{u}$%
. Also for each $\mathbf{u}$ in $\mathbb{R}^{n}$, let 
\begin{equation*}
S_{\mathbf{u}}=\{\mathbf{y\in }\mathbb{R}^{n};~n(\sigma ^{2}-\epsilon
_{n})\leq \left\Vert \mathbf{y}-\mathbf{u}\right\Vert _{2}\leq n(\sigma
^{2}+\epsilon _{n})\text{ }\}\text{ , }
\end{equation*}%
where $\ \epsilon _{n}=n^{-1/2}\log n$.
\end{Notation}

Given $S^{n-1}$, the unit sphere in $\mathbb{R}^n$, a \textbf{$\theta$%
-spherical code} is a finite subset $A\subseteq S^{n-1}$ such that no two
distinct points $\mathbf{x}, \mathbf{y} \in A$ are at an angular distance
less than $\theta$. In other words, no two points subtend an angle less than $%
\theta$ at the origin. For each $0 < \theta \leq \pi$, let $M(n, \theta)$ be
the largest cardinality of the $\theta$-spherical codes $A\subseteq S^{n-1}$%
. An important case of the spherical codes problem is the kissing number
problem, where $\theta = \pi/3$. The kissing number problem
equivalently asks for the maximum number of non-overlapping spheres that can touch
another sphere of the same size in some $n$-dimensional space.

According to our notations and definitions, for each point $\mathbf{o}$ in $%
\mathbb{R}^n$ and each radius $r$, there are at most $M(n, \theta)$ points
on the surface of ball $S(\mathbf{o},r)$, where no two points subtend an
angle less than $\theta$ at the origin $\mathbf{o}$. Independent of each
other, Chabauty \cite{Chabauty1953}, Shannon \cite{Shannon1959}, and Wyner \cite{Wyner1965} proved that 
\begin{equation*}
M(n, \theta)\geq (1-o(1))ns^{-1}_{n}(\theta)\log(\frac{\sin(\theta)}{\sqrt{2}%
\sin(\theta/2)}),
\end{equation*}
where 
\begin{equation*}
s_{n}(\theta)= (1+o(1))\frac{\sin^{n-1}(\theta)}{\sqrt{2\pi n}\cos(\theta)}.
\end{equation*}

In the sequel, by $A_{\mathbf{o}}(r,\theta)$ we mean a set of exactly $M(n,
\theta)$ points on the surface of ball $S(\mathbf{o},r)$, where no two
points subtend an angle less than $\theta$ at the origin $\mathbf{o}$. A set 
$A_{\mathbf{o}}(r,\theta)$ is called a \textbf{Galaxy of depth 1 with center 
$\mathbf{o}$} and is denoted by $A^{1}_{\mathbf{o}}(r,\theta)$. Let $k$ be a
natural number. A \textbf{Galaxy of depth 2 with center $\mathbf{o}$} is the
union of all $A^{1}_{\mathbf{u}}(r,\theta)$, where $\mathbf{u}\in A_{\mathbf{%
o}}(kr,\theta) $ and is denoted by $A^{2}_{\mathbf{o}}(r,\theta)$. Actually 
\begin{equation*}
A^{2}_{\mathbf{o}}(r,\theta)=\bigcup_{\mathbf{u}\in A_{\mathbf{o}%
}(kr,\theta)} A^{1}_{\mathbf{u}}(r,\theta).
\end{equation*}
\begin{figure}
    \centering
   \includegraphics[width=3cm]{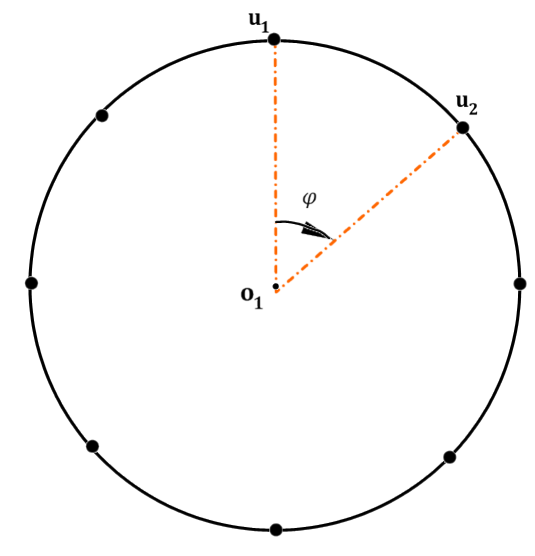} %
\includegraphics[height=9cm,width=9cm]{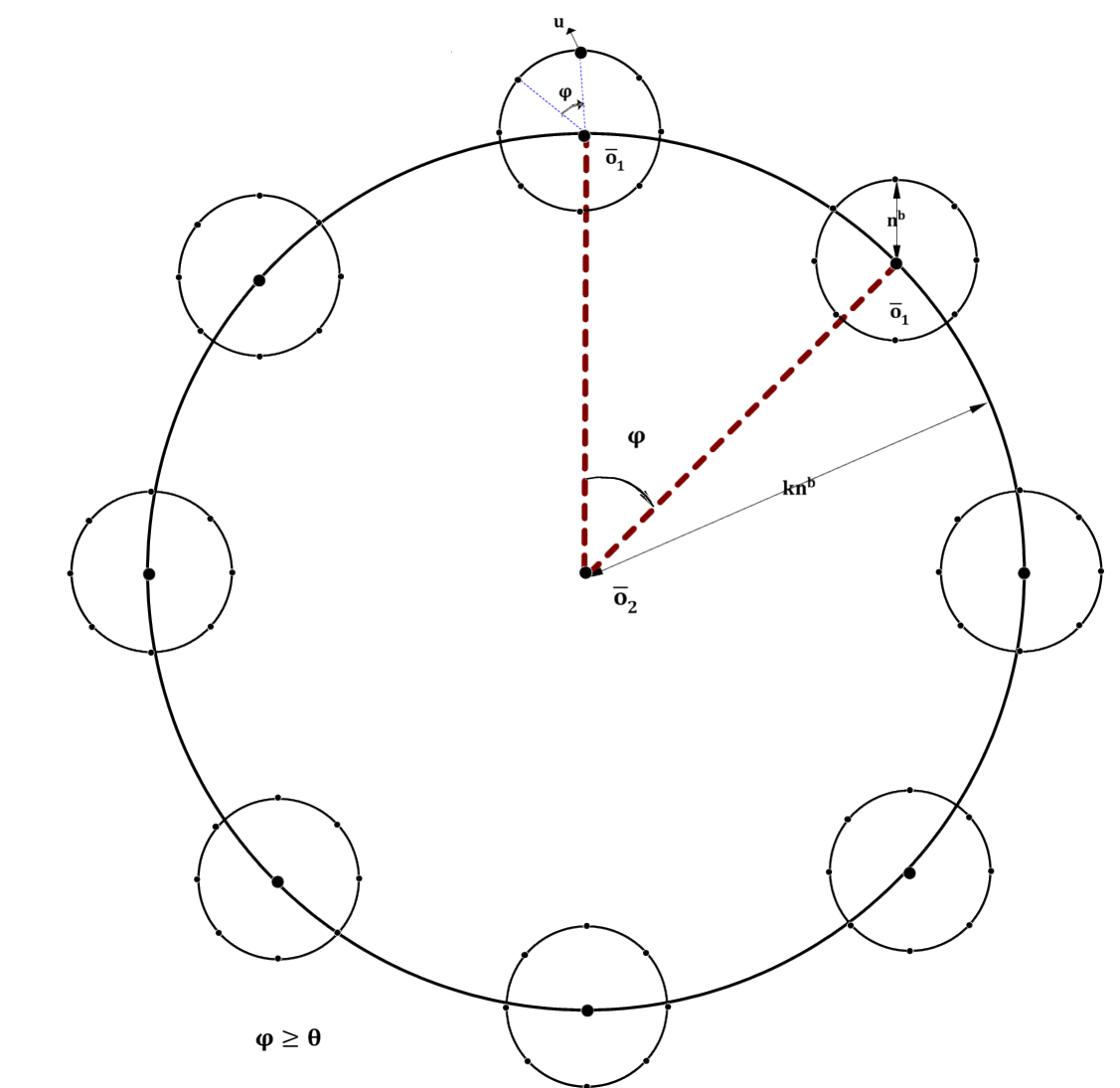}
    \caption{Illustration of a Galaxy of depth 1 and 2,
when $M(n,\protect\theta)=8$}\label{galaxy12}
\end{figure}

An illustration of a Galaxy of depth 1 and 2 is given in Fig.~\ref{galaxy12} 

\begin{figure}
    \centering
    \includegraphics[width=10cm]{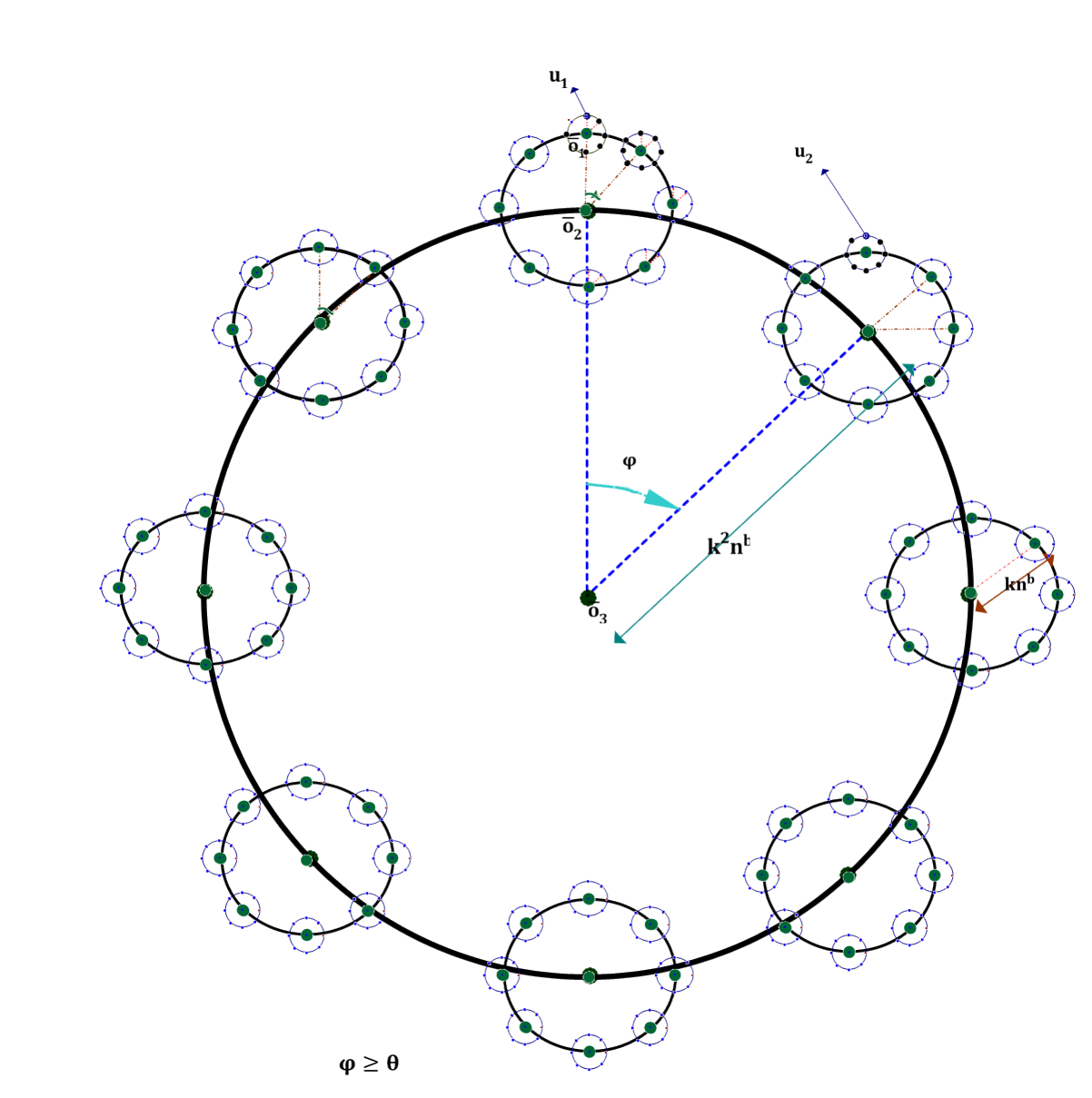}
    \caption{Illustration of a Galaxy of depth 3, when $%
M(n,\protect\theta)=8$}
    \label{depth3}
\end{figure}

A \textbf{Galaxy of depth 3 with center $\mathbf{o}$} is the union of all $%
A^{2}_{\mathbf{u}}(r,\theta)$, where $\mathbf{u}\in A_{\mathbf{o}%
}(k^{2}r,\theta)$ and is denoted by $A^{3}_{\mathbf{o}}(r,\theta)$. More
precisely 
\begin{equation*}
A^{3}_{\mathbf{o}}(r,\theta)=\bigcup_{\mathbf{u}\in A_{\mathbf{o}%
}(k^{2}r,\theta)} A^{2}_{\mathbf{u}}(r,\theta).
\end{equation*}
An illustration of a Galaxy of depth 3, when $%
M(n,\protect\theta)=8$ is given in Fig.~\ref{depth3}.

With the similar way, a \textbf{Galaxy of depth $t$ with center $\mathbf{o}$}
is denoted by $A^{t}_{\mathbf{o}}(r,\theta)$ and is defined by 
\begin{equation*}
A^{t}_{\mathbf{o}}(r,\theta)=\bigcup_{\mathbf{u}\in A_{\mathbf{o}%
}(k^{t-1}r,\theta)} A^{t-1}_{\mathbf{u}}(r,\theta).
\end{equation*}
Clearly for each $t$ we have 
\begin{equation*}
|A^{t}_{\mathbf{o}}(r,\theta)|=M(n, \theta)^{t}.
\end{equation*}
Now we are ready to define the \textbf{Galaxy Deterministic Identification
Codes}. We assume power constraint on codewords $\mathbf{u}$, that means $%
\left\Vert \mathbf{u}\right\Vert ^{2}\leq nP$, for given positive real
number $P$. For very small real number $b\in \mathbb{R}$ and natural number $%
k$ assume that $\overline{t}=\lceil\frac{\log(\frac{n^{1/4}}{n^{b}})}{\log k}%
\rceil=\lceil\frac{(1/4-b)\log n}{\log k}\rceil$, $r=n^{b}$ and $\{\mathbf{o}%
_{i},i\in \mathcal{M}\}$ are points in $\mathbb{R}^{n}$ such that $%
\left\Vert \mathbf{o}_{i}\right\Vert _{2}^{2}\leq nP$, $\left\Vert \mathbf{o}%
_{i}-\mathbf{o}_{j}\right\Vert _{2}\geq 2n^{b+1/4}$ and packing is saturated
($|\mathcal{M}| $ is maximum). Now consider a DI code $(\mathcal{U},\mathcal{%
D})_{n}$, where 
\begin{equation*}
\mathcal{U}=\bigcup_{i\in \mathcal{M}}A^{\overline{t}}_{\mathbf{o}%
_{i}}(r,\theta),
\end{equation*}
for $\overline{t}=\lceil\frac{(1/4-b)\log n}{\log k }\rceil$ and $r=n^{b}$.
For each $\mathbf{u}\in \mathcal{U}$ there is an $l\in \mathcal{M}$ such
that $\mathbf{u}\in A^{\overline{t}}_{\mathbf{o}_{l}}(r,\theta)$. Now let $%
\bar{\mathbf{o}}_{\overline{t}}=\mathbf{o}_{l}$. According to our
assumptions there is a point $\bar{\mathbf{o}}_{\overline{t}-1}\in A_{\bar{%
\mathbf{o}}_{\overline{t}}}(k^{\overline{t}-1}r,\theta)$ such that $\mathbf{u%
}\in A^{\overline{t}-1}_{\bar{\mathbf{o}}_{\overline{t}-1}}(r,\theta)$.
Again since $\mathbf{u}\in A^{\overline{t}-1}_{\bar{\mathbf{o}}_{\overline{t}%
-1}}(r,\theta)$, there is a point $\bar{\mathbf{o}}_{\overline{t}-2}\in A_{%
\bar{\mathbf{o}}_{\overline{t}-1}}(k^{\overline{t}-2}r,\theta)$ such that $%
\mathbf{u}\in A^{\overline{t}-2}_{\bar{\mathbf{o}}_{\overline{t}%
-2}}(r,\theta)$. With the same argument, we can see that there is a sequence $%
\{\bar{\mathbf{o}}_{i}\}_{i=1}^{\overline{t}}$ of points such that $\mathbf{u%
}\in \bigcap_{i=1}^{\overline{t}} A^{i}_{\bar{\mathbf{o}}_{i}}(r,\theta)$
and $\bar{\mathbf{o}}_{i}\in A_{\bar{\mathbf{o}}_{i+1}}(k^{i}r,\theta)$ for
each $1\leq i\leq \overline{t}-1$. Now consider 
\begin{equation*}
Q_{\mathbf{u}}=\bigcap_{i=1}^{\overline{t}} P_{\bar{\mathbf{o}}_{i},\mathbf{u%
}},
\end{equation*}
where 
\begin{equation*}
P_{\mathbf{o}_{i},\mathbf{u}}=\{\mathbf{y}\in \mathbb{R}^n: \|\mathbf{u}%
-Proj_{\mathbf{o}_{i}\mathbf{u}}(\mathbf{y})\|_{2}\leq \sigma \log n \}.
\end{equation*}
Now let 
\begin{equation*}
D_{\mathbf{u}}=S_{\mathbf{u}}\cap Q_{\mathbf{u}}
\end{equation*}
be the decoding set corresponding to the codeword $\mathbf{u}\in \mathcal{U}$%
. We will show that $(\mathcal{U},\mathcal{D})_n$ is an identification code
over Gaussian channels. We call this code a \textbf{Galaxy code} and denote
it by $\mathrm{G}_{n}(\theta,b,k)$.

\section{Properties of Galaxy codes}\label{properties}

\subsection{ Number of codewords}

Let $N=|\mathcal{U}|$ be the number of our codewords of a Galaxy code $%
\mathrm{G}_{n}(\theta,b,k)$. We know that $|A^{\overline{t}}_{\mathbf{o}%
_{i}}(r,\theta)|=M(n, \theta)^{\overline{t}}$. On the other hand,

\begin{eqnarray*}
M(n, \theta)&\geq& (1-o(1))ns^{-1}_{n}(\theta)\log(\frac{\sin(\theta)}{\sqrt{2}%
\sin(\theta/2)})\\
&\simeq & \frac{(1-o(1))n\sqrt{2\pi n}\cos(\theta)\log(\frac{\sin(\theta)}{\sqrt{2}%
\sin(\theta/2)})}{(1+o(1))\sin^{n-1}(\theta)}\\
&\geq& \frac{1}{\sin^{n}(\theta)}.
\end{eqnarray*}

Therefore, we have 

\begin{eqnarray*}
 N=|\mathcal{M}||A^{\overline{t}}_{\mathbf{o}_{i}}(r,\theta)|=|\mathcal{M}%
|M(n, \theta)^{\overline{t}}\geq |\mathcal{M}|\frac{1}{\sin^{n\overline{t}%
}(\theta)}.
\end{eqnarray*}

Now we show the following claim about the number of codewords.

\bigskip

\textbf{Claim 1: }If \ $|\mathcal{M}|=2^{nR _{1}\log n}$\bigskip , then%
\begin{align*}
R_{1} &\geq  \frac{\log \sqrt{nP}-1}{\log n}-(b+1/4)\\ 
R_{1}   &\leq \frac{\log (\sqrt{nP}%
+n^{b+1/4})}{\log n}-(b+1/4).  \tag{1}
\end{align*}

\textbf{Proof: } Let $\mathbf{o}=(0,0,\cdots,0)$. Clearly $|\mathcal{M}|$
balls $S(\mathbf{o}_{i},n^{b+1/4})$ are pairwise disjoint. So 
\begin{eqnarray*}
|\mathcal{M}|Vol(S(\mathbf{o}_{1},n^{b+1/4})) &=&\sum\limits_{i=1}^{|%
\mathcal{M}|}Vol(S(\mathbf{o}_{i},n^{b+1/4})) \\
&\leq &Vol(S(\mathbf{o},\sqrt{nP}+n^{b+1/4})).
\end{eqnarray*}

Therefore 
\begin{equation*}
|\mathcal{M}|\leq \frac{Vol(S(\mathbf{o},\sqrt{nP}+n^{b+1/4}))}{Vol(S(%
\mathbf{u}_{1},n^{b+1/4}))}=(\frac{\sqrt{nP}+n^{b+1/4}}{n^{b+1/4}})^{n},
\end{equation*}

and so 
\begin{equation*}
n\log n\text{ }R _{1}\leq n(\log (\sqrt{nP}+n^{b+1/4})-(b+1/4)\log n),
\end{equation*}%
which implies that 
\begin{equation*}
R _{1}\leq \frac{\log (\sqrt{nP}+n^{b+1/4})}{\log n}-(b+1/4).
\end{equation*}

On the other hand, by maximality of $|\mathcal{M}|$, we conclude that $$S(%
\mathbf{o},\sqrt{nP})\subseteq \bigcup\limits_{i=1}^{|\mathcal{M}|}S(\mathbf{%
o}_{i},2n^{b+1/4})$$ and then $Vol(S(\mathbf{o},\sqrt{nP}))\leq 2^{n}.|%
\mathcal{M}|.Vol(S(\mathbf{o}_{1},n^{b+1/4})).$ This implies that%
\begin{equation*}
|\mathcal{M}|\geq 2^{-n}\frac{Vol(S(\mathbf{o},\sqrt{nP}))}{Vol(S(\mathbf{o}%
_{1},n^{b+1/4}))}=2^{-n}(\frac{\sqrt{nP}}{n^{b+1/4}})^{n}.
\end{equation*}

Therefore,%
\begin{equation*}
n\log n\text{ }R _{1}\geq -n+n(\log \sqrt{nP}-(b+1/4)\log n),
\end{equation*}%
and then%
\begin{equation*}
R _{1}\geq -\frac{1}{\log n}+\frac{\log \sqrt{nP}}{\log n}-(b+1/4)=\frac{%
\log \sqrt{nP}-1}{\log n}-(b+1/4).
\end{equation*}

Now using Claim 1, the equality $\overline{t}=\lceil\frac{(1/4-b)\log n }{%
\log k }\rceil$ and the fact that 
\begin{equation*}
N=|\mathcal{M}||A^{\overline{t}}_{o_{i}}(r,\theta)|=|\mathcal{M}|M(n,
\theta)^{\overline{t}}\geq |\mathcal{M}|\frac{1}{\sin^{n\overline{t}}(\theta)%
},
\end{equation*}
we conclude with the following result.

\begin{Lemma}
\label{1}
If \ $N=2^{nR\log n}$\bigskip , then 
\begin{equation*}
\frac{\log \sqrt{nP}-1}{\log n}-(-b+1/4)\frac{\log(\sin(\theta))}{\log k }%
-(b+1/4)\leq R.
\end{equation*}
\end{Lemma}

\subsection{Distances between codewords}

In this section, we give some results about the distances between codewords
of Galaxy codes.

\begin{Theorem}
\label{cond1} For $\mathbf{u}\in A^{t}_{\mathbf{o}}(r,\theta)$, we have 
\begin{equation*}
r\frac{k^{t}-2k^{t-1}+1}{k-1}\leq\|\mathbf{u}-\mathbf{o}\|_{2}\leq r\frac{%
k^{t}-1}{k-1}.
\end{equation*}
\end{Theorem}

\textbf{Proof:} Let $\mathbf{o}_{t}=\mathbf{o}$. If $\mathbf{u}\in A^{t}_{\mathbf{o}%
}(r,\theta)=A^{t}_{\mathbf{o}_{t}}(r,\theta)$, then according to our
assumptions there is a point $\mathbf{o}_{t-1}\in A_{\mathbf{o}%
_{t}}(k^{t-1}r,\theta)$ such that $\mathbf{u}\in A^{t-1}_{\mathbf{o}%
_{t-1}}(r,\theta)$. Again since $\mathbf{u}\in A^{t-1}_{\mathbf{o}%
_{t-1}}(r,\theta)$, there is a point $\mathbf{o}_{t-2}\in A_{\mathbf{o}%
_{t-1}}(k^{t-2}r,\theta)$ such that $\mathbf{u}\in A^{t-2}_{\mathbf{o}%
_{t-2}}(r,\theta)$. With the same argument, we can see that there is a
sequence $\{\mathbf{o}_{i}\}_{i=1}^{t}$ of points such that $\mathbf{u}\in
\bigcap_{i=1}^{t} A^{i}_{\mathbf{o}_{i}}(r,\theta)$ and $\mathbf{o}_{i}\in
A_{\mathbf{o}_{i+1}}(k^{i}r,\theta)$ for each $1\leq i\leq t-1$. Now clearly
we have 
\begin{align*}
 \|\mathbf{u}-\mathbf{o}\|_{2} & =\|\mathbf{u}-\mathbf{o}_{t}\|_{2}\\
 & \leq \|%
\mathbf{u}-\mathbf{o}_{1}\|_{2}+\sum_{i=1}^{t-1}\|\mathbf{o}_{i}-\mathbf{o}%
_{i+1}\|_{2}\\
& \leq r+\sum_{i=1}^{t-1}k^{i}r=r\frac{k^{t}-1}{k-1}.    
\end{align*}

On the other hand, we have 
\begin{align*}
k^{t-1}r & =\|\mathbf{o}_{t}-\mathbf{o}_{t-1}\|_{2}\\
& \leq \|\mathbf{o}_{t}-%
\mathbf{u}\|_{2}+\|\mathbf{u}-\mathbf{o}_{1}\|_{2}+\sum_{i=1}^{t-2}\|\mathbf{%
o}_{i}-\mathbf{o}_{i+1}\|_{2}\\
& \leq \|\mathbf{u}-\mathbf{o}\|_{2}+r+%
\sum_{i=1}^{t-2}k^{i}r.
\end{align*}
Therefore 
\begin{equation*}
r(k^{t-1}-\sum_{i=0}^{t-2}k^{i})=r\frac{k^{t}-2k^{t-1}+1}{k-1}\leq\|\mathbf{%
u}-\mathbf{o}\|_{2},
\end{equation*}
and the proof is completed.

For given two codewords $\mathbf{u}_{1}, \mathbf{u}_{2}\in A^{t}_{\mathbf{o}%
}(r,\theta)$, according to our definitions there are two sequences $\{%
\mathbf{o}_{1,i}\}_{i=1}^{t}$ and $\{\mathbf{o}_{2,i}\}_{i=1}^{t}$ of points
such that for $j=1,2$, we have $\mathbf{u}_{j}\in \bigcap_{i=1}^{t} A^{i}_{%
\mathbf{o}_{j,i}}(r,\theta)$ and $\mathbf{o}_{j,i}\in A_{\mathbf{o}%
_{j,i+1}}(k^{i}r,\theta)$ for each $1\leq i\leq t-1$. Note that $\mathbf{o}%
_{1,t}=\mathbf{o}_{2,t}=\mathbf{o}$ and so $\{\mathbf{o}_{1,i}\}_{i=1}^{t}%
\cap \{\mathbf{o}_{2,i}\}_{i=1}^{t}$ is nonempty. Now assume that $t(\mathbf{%
u}_{1}, \mathbf{u}_{2})$ is the minimum value of $l$ for which $\mathbf{u}%
_{1}, \mathbf{u}_{2}\in A^{l}_{\bar{\mathbf{o}}}(r,\theta)$, for some $\bar{%
\mathbf{o}}\in \{\mathbf{o}_{1,i}\}_{i=1}^{t}\cap \{\mathbf{o}%
_{2,i}\}_{i=1}^{t}$.

\begin{Theorem}
\label{cond2} For given codewords $\mathbf{u}_{1}, \mathbf{u}_{2}\in A^{t}_{%
\mathbf{o}}(r,\theta)$ assume that $t(\mathbf{u}_{1}, \mathbf{u}_{2})$ is
the minimum value of $l$ for which $\mathbf{u}_{1}, \mathbf{u}_{2}\in A^{l}_{%
\bar{\mathbf{o}}}(r,\theta)$, for some $\bar{\mathbf{o}}\in \{\mathbf{o}%
_{1,i}\}_{i=1}^{t}\cap \{\mathbf{o}_{2,i}\}_{i=1}^{t}$. Then we have 
\begin{equation*}
2r(k^{t(\mathbf{u}_{1}, \mathbf{u}_{2})-1}\sin(\theta/2)-\frac{k^{t(\mathbf{u%
}_{1}, \mathbf{u}_{2})-1}-1}{k-1})\leq\|\mathbf{u}_{1}-\mathbf{u}_{2}\|_{2}.
\end{equation*}
\end{Theorem}

\textbf{Proof: } Since $t(\mathbf{u}_{1}, \mathbf{u}_{2})$ is the minimum value of $l$
for which $\mathbf{u}_{1}, \mathbf{u}_{2}\in A^{l}_{\bar{\mathbf{o}}%
}(r,\theta)$, for some $\bar{\mathbf{o}}\in \{\mathbf{o}_{1,i}\}_{i=1}^{t}%
\cap \{\mathbf{o}_{2,i}\}_{i=1}^{t}$, according to our definitions we have $%
\bar{\mathbf{o}}=\mathbf{o}_{1,t(\mathbf{u}_{1}, \mathbf{u}_{2})}=\mathbf{o}%
_{2,t(\mathbf{u}_{1}, \mathbf{u}_{2})}$ and $o_{1,t(\mathbf{u}_{1}, \mathbf{u%
}_{2})-1},o_{2,t(\mathbf{u}_{1}, \mathbf{u}_{2})-1}\in A_{\bar{\mathbf{o}}%
}(k^{t(\mathbf{u}_{1}, \mathbf{u}_{2})-1}r,\theta)$. Therefore 
\begin{align*}
\|\mathbf{o}_{1,t(\mathbf{u}_{1}, \mathbf{u}_{2})-1}-\mathbf{o}_{2,t(\mathbf{%
u}_{1}, \mathbf{u}_{2})-1}\|_{2} & = 2k^{t(\mathbf{u}_{1}, \mathbf{u}%
_{2})-1}r\sin(\alpha/2)\\
& \geq 2k^{t(\mathbf{u}_{1}, \mathbf{u}%
_{2})-1}r\sin(\theta/2),
\end{align*}
where $\alpha$ is the angle between two points $\mathbf{o}_{1,t(\mathbf{u}%
_{1}, \mathbf{u}_{2})-1}$ and $\mathbf{o}_{2,t(\mathbf{u}_{1}, \mathbf{u}%
_{2})-1}$ at the origin $\bar{\mathbf{o}}$. According to our definitions and
code construction we have $\pi\geq \alpha\geq \theta$ and so $%
\sin(\alpha/2)\geq \sin(\theta/2)$. Now clearly we have 
\begin{align*}
&\|\mathbf{u}_{1}-\mathbf{u}_{2}\|_{2} \geq \\
&\|\mathbf{o}_{1,t(\mathbf{u}_{1}, 
\mathbf{u}_{2})-1}-\mathbf{o}_{2,t(\mathbf{u}_{1}, \mathbf{u}%
_{2})-1}\|_{2}\\
&-\|\mathbf{o}_{1,t(\mathbf{u}_{1}, \mathbf{u}_{2})-1}-\mathbf{u%
}_{1}\|_{2}-\|\mathbf{o}_{2,t(\mathbf{u}_{1}, \mathbf{u}_{2})-1}-\mathbf{u}%
_{2}\|_{2},
\end{align*}
On the other hand, for $i=1,2$ we have 
\begin{equation*}
\mathbf{u}_{i}\in A^{t(\mathbf{u}_{1}, \mathbf{u}_{2})-1}_{\mathbf{o}_{i,t(%
\mathbf{u}_{1}, \mathbf{u}_{2})-1}}(r,\theta).
\end{equation*}
Therefore using Theorem~\ref{cond1} we have 
\begin{equation*}
\|\mathbf{o}_{1,t(\mathbf{u}_{1}, \mathbf{u}_{2})-1}-\mathbf{u}%
_{1}\|_{2}\leq r\frac{k^{t(\mathbf{u}_{1}, \mathbf{u}_{2})-1}-1}{k-1},
\end{equation*}
and 
\begin{equation*}
\|\mathbf{o}_{2,t(\mathbf{u}_{1}, \mathbf{u}_{2})-1}-\mathbf{u}%
_{2}\|_{2}\leq r\frac{k^{t(\mathbf{u}_{1}, \mathbf{u}_{2})-1}-1}{k-1},
\end{equation*}
and so 
\begin{align*}
\|\mathbf{u}_{1}-\mathbf{u}_{2}\|_{2} \geq &\|\mathbf{o}_{1,t(\mathbf{u}%
_{1}, \mathbf{u}_{2})-1}-\mathbf{o}_{2,t(\mathbf{u}_{1}, \mathbf{u}%
_{2})-1}\|_{2}-\\
& \|\mathbf{o}_{1,t(\mathbf{u}_{1}, \mathbf{u}_{2})-1}-\mathbf{u%
}_{1}\|_{2}-\|\mathbf{o}_{2,t(\mathbf{u}_{1}, \mathbf{u}_{2})-1}-\mathbf{u}%
_{2}\|_{2} \\
\geq  & 2k^{t(\mathbf{u}_{1}, \mathbf{u}_{2})-1}r\sin(\theta/2)-2r\frac{k^{t(%
\mathbf{u}_{1}, \mathbf{u}_{2})-1}-1}{k-1},
\end{align*}
and the proof is completed. 

Consider two points $\mathbf{o}$ and $\mathbf{u}$ in $\mathbb{R}^n$. For
each point $\mathbf{y}\in \mathbb{R}^n$ the projection of $\mathbf{y}$ on
the line from two points $\mathbf{o}$ and $\mathbf{u}$ is denoted by $Proj_{%
\mathbf{o}\mathbf{u}}(\mathbf{y})$. Let 
\begin{equation*}
P_{\mathbf{o},\mathbf{u}}=\{\mathbf{y}\in R^{n}: \|\mathbf{u}-Proj_{\mathbf{o%
}\mathbf{u}}(\mathbf{y})\|_{2}\leq \sigma \log n \}.
\end{equation*}%
.

\begin{Lemma}
\label{condglog} For given codewords $\mathbf{u}_{1}, \mathbf{u}_{2}\in
A^{t}_{\mathbf{o}}(r,\theta)$ assume that $t(\mathbf{u}_{1}, \mathbf{u}_{2})$
is the minimum value of $l$ for which $\mathbf{u}_{1}, \mathbf{u}_{2}\in
A^{l}_{\bar{\mathbf{o}}}(r,\theta)$, for some $\bar{\mathbf{o}}\in \{\mathbf{%
o}_{1,i}\}_{i=1}^{t}\cap \{\mathbf{o}_{2,i}\}_{i=1}^{t}$. Then $P(P_{\bar{%
\mathbf{o}},\mathbf{u}_{1}}|\mathbf{u}_{2})\leq 2\Phi (-\log n)$ when 
\begin{equation*}
\frac{\|\mathbf{u}_{1}-\bar{\mathbf{o}}\|^{2}_{2}+\|\mathbf{u}_{1}-\mathbf{u}%
_{2}\|^{2}_{2}-\|\mathbf{u}_{2}-\bar{\mathbf{o}}\|^{2}_{2}}{2\|\mathbf{u}%
_{1}-\bar{\mathbf{o}}\|_{2}}\geq 2\sigma\log n.
\end{equation*}
\end{Lemma}

\textbf{Proof:} Assume that the projection of $\mathbf{u}_{2}$ on the line
from two points $\bar{\mathbf{o}}$ and $\mathbf{u}_{1}$ is 
\begin{equation*}
\mathbf{w}=Proj_{\bar{\mathbf{o}}\mathbf{u}_{1}}(\mathbf{u}_{2}).
\end{equation*}
Clearly 
\begin{equation*}
\|\mathbf{w}-\bar{\mathbf{o}}\|^{2}_{2}+\|\mathbf{w}-\mathbf{u}%
_{2}\|^{2}_{2}=\|\mathbf{u}_{2}-\bar{\mathbf{o}}\|^{2}_{2},
\end{equation*}
and 
\begin{equation*}
\|\mathbf{w}-\mathbf{u}_{1}\|^{2}_{2}+\|\mathbf{w}-\mathbf{u}%
_{2}\|^{2}_{2}=\|\mathbf{u}_{2}-\mathbf{u}_{1}\|^{2}_{2},
\end{equation*}
and so 
\begin{equation*}
\|\mathbf{w}-\bar{\mathbf{o}}\|^{2}_{2}-\|\mathbf{w}-\mathbf{u}%
_{1}\|^{2}_{2}=\|\mathbf{u}_{2}-\bar{\mathbf{o}}\|^{2}_{2}-\|\mathbf{u}_{2}-%
\mathbf{u}_{1}\|^{2}_{2}.
\end{equation*}
On the other hand, 
\begin{equation*}
\|\mathbf{w}-\bar{\mathbf{o}}\|_{2}+\|\mathbf{w}-\mathbf{u}_{1}\|_{2}=\|%
\mathbf{u}_{1}-\bar{\mathbf{o}}\|_{2}.
\end{equation*}
Therefore, 
\begin{equation*}
\|\mathbf{w}-\mathbf{u}_{1}\|_{2}=\frac{\|\mathbf{u}_{1}-\bar{\mathbf{o}}%
\|^{2}_{2}+\|\mathbf{u}_{1}-\mathbf{u}_{2}\|^{2}_{2}-\|\mathbf{u}_{2}-\bar{%
\mathbf{o}}\|^{2}_{2}}{2\|\mathbf{u}_{1}-\bar{\mathbf{o}}\|_{2}}.
\end{equation*}
According to our assumptions we conclude that $\|\mathbf{w}-\mathbf{u}%
_{1}\|_{2}\geq 2\sigma\log n$. Therefore

\begin{eqnarray*}
&& P(P_{\bar{\mathbf{o}},\mathbf{u}_{1}} |\mathbf{u}_{2})\\ &=&P( \|\mathbf{u}%
_1-Proj_{\bar{\mathbf{o}}\mathbf{u}_1}(\mathbf{Y})\|_{2}\leq \sigma \log n | 
\mathbf{u}_{2}) \\
&=&P(\|\mathbf{u}_1-Proj_{\bar{\mathbf{o}}\mathbf{u}_1}(\mathbf{\mathbf{u}%
_{2}+\sigma\mathbf{Z}})\|_{2}\leq \sigma\log n) \\
&=&P(\|\mathbf{u}_1-(Proj_{\bar{\mathbf{o}}\mathbf{u}_1}(\mathbf{u}%
_{2})+Proj_{\bar{\mathbf{o}}\mathbf{u}_1} (\sigma{\vec{\mathbf{Z}}}%
)\|_{2}\leq \sigma \log n) \\
&=&P(\|Proj_{\bar{\mathbf{o}}\mathbf{u}_1}(\sigma\vec{\mathbf{Z}})+\mathbf{w}%
- \mathbf{u}_1 \|_{2}\leq \sigma \log n) \\
&\leq&P(-\|Proj_{\bar{\mathbf{o}}\mathbf{u}_1}(\sigma\vec{\mathbf{Z}})
\|_{2}+\|\mathbf{w}- \mathbf{u}_1 \|_2\leq \sigma \log n) \\
&\leq&P(\|Proj_{\bar{\mathbf{o}}\mathbf{u}_1}(\vec{\mathbf{Z}}) \|_{2}\geq
\log n) \\
&=&2\Phi (-\log n),
\end{eqnarray*}
where the last inequality is held by Appendix B.

Note that according to Notations \ref{Note1}, we have 
$Proj_{\bar{\mathbf{o}}_{i}\mathbf{u_{1}}}(\mathbf{u_{2}}+\sigma \mathbf{Z}%
)=Proj_{\bar{\mathbf{o}}_{i}\mathbf{u_{1}}}(\mathbf{u_{2}})+Proj_{\bar{%
\mathbf{o}}_{i}\mathbf{u}}(\sigma \vec{\mathbf{Z}})$, where $Proj_{\bar{%
\mathbf{o}}_{i}\mathbf{u_{1}}}(\mathbf{u_{2}})$ is the projection of the
point $\mathbf{u_{2}}$ on the line $\bar{\mathbf{o}}_{i}\mathbf{u_{1}}$ and $%
Proj_{\bar{\mathbf{o}}_{i}\mathbf{u_{1}}}(\sigma \vec{\mathbf{Z}})$ is the
projection of the vector $\sigma \vec{\mathbf{Z}}$ on the vector $\overset{%
\longrightarrow }{\mathbf{\bar{o}}_{i}\mathbf{u}_{1}}$. If we consider $Proj_{\bar{\mathbf{o}}_{i}\mathbf{u_{1}%
}}(\mathbf{u_{2}})+Proj_{\bar{\mathbf{o}}_{i}\mathbf{u_{1}}}(\sigma \vec{%
\mathbf{Z}})$ as a summation between two $n$-tuples in $\mathbb{R}^{n}$, the
result is the coordinate of the point $Proj_{\bar{\mathbf{o}}_{i}\mathbf{%
u_{1}}}(\mathbf{u_{2}}+\sigma \mathbf{Z})$ (the projection of the point with
coordinate $\mathbf{u_{2}}+\sigma \mathbf{Z}$ on the line $\bar{\mathbf{o}}%
_{i}\mathbf{u_{1}}$).

\begin{Theorem}
For given codewords $\mathbf{u}_{1}, \mathbf{u}_{2}\in A^{t}_{\mathbf{o}%
}(r,\theta)$, assume that $t(\mathbf{u}_{1}, \mathbf{u}_{2})$ is the minimum
value of $l$ for which $\mathbf{u}_{1}, \mathbf{u}_{2}\in A^{l}_{\bar{%
\mathbf{o}}}(r,\theta)$ for some $\bar{\mathbf{o}}\in \{\mathbf{o}%
_{1,i}\}_{i=1}^{t}\cap \{\mathbf{o}_{2,i}\}_{i=1}^{t}$. If $(\sin(\theta/2)-%
\frac{1}{k-1})^{2}>\frac{2}{k-1}$, then $P(P_{\bar{\mathbf{o}},\mathbf{u}%
_{1}}|\mathbf{u}_{2})\leq 2\Phi (-\log n).$
\end{Theorem}

\textbf{Proof:} According to our assumptions, $\mathbf{u}_{1}, \mathbf{u}_{2}\in
A^{t(\mathbf{u}_{1}, \mathbf{u}_{2})}_{\bar{\mathbf{o}}}(r,\theta),$ and so
using Theorem \ref{cond1}, we have 
\begin{equation*}
r\frac{k^{t(\mathbf{u}_{1}, \mathbf{u}_{2})}-2k^{t(\mathbf{u}_{1}, \mathbf{u%
}_{2})-1}+1}{k-1}\leq\|\bar{\mathbf{o}}-\mathbf{u}_{1}\|_{2}\leq r\frac{%
k^{t(\mathbf{u}_{1}, \mathbf{u}_{2})}-1}{k-1},
\end{equation*}
and 
\begin{equation*}
\|\bar{\mathbf{o}}-\mathbf{u}_{2}\|_{2}\leq r\frac{k^{t(\mathbf{u}_{1}, 
\mathbf{u}_{2})}-1}{k-1}.
\end{equation*}%

On the other hand, using Theorem \ref{cond2}, we have

\begin{equation*}
2r(k^{t(\mathbf{u}_{1}, \mathbf{u}_{2})-1}\sin(\theta/2)-\frac{k^{t(\mathbf{u%
}_{1}, \mathbf{u}_{2})-1}-1}{k-1})\leq\|\mathbf{u}_{1}-\mathbf{u}_{2}\|_{2}.
\end{equation*}

Therefore,
\begin{equation*}
\frac{\|\mathbf{u}_{1}-\bar{\mathbf{o}}\|^{2}_{2}+\|\mathbf{u}_{1}-\mathbf{u}%
_{2}\|^{2}_{2}-\|\mathbf{u}_{2}-\bar{\mathbf{o}}\|^{2}_{2}}{2\|\mathbf{u}%
_{1}-\bar{\mathbf{o}}\|_{2}}
\end{equation*}%
\begin{equation*}
\geq 4r(k-1)k^{2(t(\mathbf{u}_{1}, \mathbf{u}_{2})-1)}\frac{(\sin(\theta/2)-%
\frac{1}{k-1})^{2}-\frac{1}{k-1}}{k^{t(\mathbf{u}_{1}, \mathbf{u}_{2})}-1}
\end{equation*}%
\begin{equation*}
\geq \frac{4rk^{2(t(\mathbf{u}_{1}, \mathbf{u}_{2})-1)}}{k^{t(\mathbf{u}%
_{1}, \mathbf{u}_{2})}-1}\geq 4r=4n^b\geq 2\log n.
\end{equation*}

Therefore using Lemma \ref{condglog}, we conclude that $P(P_{\bar{\mathbf{o}}%
,\mathbf{u}_{1}}|\mathbf{u}_{2})\leq 2\Phi (-\log n),$ and so we are done.

\begin{Corollary}
\label{error2} If $(\sin(\theta/2)-\frac{1}{k-1})^{2}>\frac{1}{k-1}$, then
for any two codewords $\mathbf{u}_{1}, \mathbf{u}_{2}$ in the same Galaxy of
depth $\overline{t}=\lceil\frac{(1/4-b)\log n }{\log k }\rceil$ (this means
that $\mathbf{u}_{1}, \mathbf{u}_{2}\in A^{\overline{t}}_{\mathbf{o}%
_{i}}(r,\theta) $ for some $i\in \mathcal{M}$), we have $P(P_{\bar{\mathbf{o}%
},\mathbf{u}_{1}}|\mathbf{u}_{2})\leq 2\Phi (-\log n)$.
\end{Corollary}

\begin{Lemma}
\label{diffgalax} 
For any two codewords $\mathbf{u}_{1}, \mathbf{u}_{2}$ in
the different Galaxies of depth $$\overline{t}=\lceil\frac{(1/4-b)\log n }{\log k }\rceil ,$$ (which means $\mathbf{u}_{1}\in A^{\overline{t}}_{%
\mathbf{o}_{i}}(r,\theta),~\mathbf{u}_{2}\in A^{\overline{t}}_{\mathbf{o}%
_{j}}(r,\theta)$ for some $i,j\in \mathcal{M}$ with $i\neq j$) we have

\begin{equation*}
\|\mathbf{u}_{1}-\mathbf{u}_{2}\|_{2}\geq \frac{n^{b+1/4}}{2}.
\end{equation*}
\end{Lemma}

Proof: We have 
\begin{equation*}
\|\mathbf{u}_{1}-\mathbf{u}_{2}\|_{2}\geq \|\mathbf{o}_{i}-\mathbf{o}%
_{j}\|_{2}-\|\mathbf{u}_{1}-\mathbf{o}_{i}\|_{2}-\|\mathbf{u}_{2}-\mathbf{o}%
_{j}\|_{2}.
\end{equation*}
On the other hand, according to our code construction we have $\|\mathbf{o}%
_{i}-\mathbf{o}_{j}\|_{2}\geq n^{b+1/4}$ and by Theorem \ref{cond1}

\begin{equation*}
\|\mathbf{u}_{1}-\mathbf{o}_{i}\|_{2}, \|\mathbf{u}_{2}-\mathbf{o}%
_{j}\|_{2}\leq r\frac{k^{\overline{t}}-1}{k-1}.
\end{equation*}

Therefore for $\overline{t}=\lceil\frac{(1/4-b)\log n}{\log k }\rceil$ we
have 
\begin{equation*}
\|\mathbf{u}_{1}-\mathbf{u}_{2}\|_{2}\geq n^{b+1/4}-2r\frac{k^{\overline{t}%
}-1}{k-1}\geq \frac{n^{b+1/4}}{2}.
\end{equation*}

\subsection{ \textbf{Error Analysis}}

\textbf{Type I probability error: }

For each $\mathbf{u}\in \mathcal{U}$, as we mentioned in Section 3, there is
a sequence $\{\bar{\mathbf{o}}_{i}\}_{i=1}^{\overline{t}}$ of points such
that $\mathbf{u}\in \bigcap_{i=1}^{\overline{t}} A^{i}_{\bar{\mathbf{o}}%
_{i}}(r,\theta)$ and $\bar{\mathbf{o}}_{i}\in A_{\bar{\mathbf{o}}%
_{i+1}}(k^{i}r,\theta)$ for each $1\leq i\leq \overline{t}-1$. Also 
\begin{equation*}
P_{\bar{\mathbf{o}}_{i},\mathbf{u}}=\{\mathbf{y}\in \mathbb{R}^n: \|\mathbf{u%
}-Proj_{\bar{\mathbf{o}}_{i}\mathbf{u}}(\mathbf{y})\|_{2}\leq \sigma \log n
\},
\end{equation*}
and $Q_{\mathbf{u}}=\bigcap_{i=1}^{\overline{t}} P_{\bar{\mathbf{o}}_{i},%
\mathbf{u}}$ and $D_{\mathbf{u}}=S_{\mathbf{u}}\cap Q_{\mathbf{u}}$.

Then, we have 
\begin{eqnarray*}
&& P(P_{\bar{\mathbf{o}}_{i},\mathbf{u}}|\mathbf{u})\\&=& P(\|\mathbf{u}-Proj_{%
\bar{\mathbf{o}}_{i}\mathbf{u}}(\mathbf{Y})\|_{2}\leq \sigma \log n|\mathbf{u%
}) \\
&=&P(\|\mathbf{u}-Proj_{\bar{\mathbf{o}}_{i}\mathbf{u}}(\mathbf{u}+\sigma 
\mathbf{Z})\|_{2}\leq \sigma \log n) \\
&=&P(\|\mathbf{u}-(Proj_{\bar{\mathbf{o}}_{i}\mathbf{u}}(\mathbf{u})+Proj_{%
\bar{\mathbf{o}}_{i}\mathbf{u}}(\sigma\vec{\mathbf{Z}}))\|_{2}\leq \sigma
\log n ) \\
&=&P(\|Proj_{\bar{\mathbf{o}}_{i}\mathbf{u}}(\sigma\vec{\mathbf{Z}}%
)\|_{2}\leq \sigma \log n ) \\
&=&P(\|Proj_{\bar{\mathbf{o}}_{i}\mathbf{u}}(\vec{\mathbf{Z}})\|_{2}\leq
\log n).
\end{eqnarray*}

Note that in the above equations according to Notations 3.1, we have 
\begin{equation*}
Proj_{\bar{\mathbf{o}}_{i}\mathbf{u}}(\mathbf{u}+\sigma \mathbf{Z})=Proj_{%
\bar{\mathbf{o}}_{i}\mathbf{u}}(\mathbf{u})+Proj_{\bar{\mathbf{o}}_{i}%
\mathbf{u}}(\sigma\vec{\mathbf{Z}}).
\end{equation*}
$Proj_{\bar{\mathbf{o}}_{i}\mathbf{u}}(\mathbf{u})$ is the projection of the
point $\mathbf{u}$ on the line $\bar{\mathbf{o}}_{i}\mathbf{u}$, which is
clearly equal to $\mathbf{u}$ and $Proj_{\bar{\mathbf{o}}_{i}\mathbf{u}%
}(\sigma\vec{\mathbf{Z}})$ is the projection of the vector $\sigma\vec{%
\mathbf{Z}}$ on the vector $\vec {\bar{\mathbf{o}}}_{i}\mathbf{u}$. As we
mentioned in Notations 3.1, if we consider $Proj_{\bar{\mathbf{o}}_{i}%
\mathbf{u}}(\mathbf{u})+Proj_{\bar{\mathbf{o}}_{i}\mathbf{u}}(\sigma\vec{%
\mathbf{Z}})$ as a summation between two $n$-tuples in $\mathbb{R}^{n}$, the
result is the coordinate of the point $Proj_{\bar{\mathbf{o}}_{i}\mathbf{u}}(%
\mathbf{u}+\sigma \mathbf{Z})$ (the projection of the point with coordinate $%
\mathbf{u}+\sigma \mathbf{Z}$ on the line $\bar{\mathbf{o}}_{i}\mathbf{u}$).
In Appendix B, it has been shown that 

\begin{equation*}
P(\|Proj_{\bar{\mathbf{o}}_{i}\mathbf{u}}(\vec{\mathbf{Z}})\|_{2}\leq \log n
)=1-2\Phi (-\log n).
\end{equation*}%

And then, 
\begin{eqnarray*}
P(P_{\bar{\mathbf{o}}_{i},\mathbf{u}}|\mathbf{u})&=& 1-2\Phi (-\log n),
\end{eqnarray*}

and so

\begin{eqnarray*}
P(\mathbb{R}^{n}-P_{\bar{\mathbf{o}}_{i},\mathbf{u}}|\mathbf{u})&=& 2\Phi
(-\log n).
\end{eqnarray*}
Therefore we have 
\begin{eqnarray*}
P(\mathbb{R}^{n}-Q_{\mathbf{u}}|\text{ }\mathbf{u})&=&P(\mathbb{R}%
^{n}-\bigcap_{i=1}^{\overline{t}} P_{\bar{\mathbf{o}}_{i},\mathbf{u}}|\text{ 
}\mathbf{u}) \\
&\leq& \sum_{i=1}^{\overline{t}}P(\mathbb{R}^{n}-P_{\bar{\mathbf{o}}_{i},%
\mathbf{u}}|\mathbf{u}) \\
&=&2\overline{t}\Phi (-\log n) \\
&=&2\lceil\frac{(1/4-b)\log n}{\log n }\rceil\Phi (-\log n)
\end{eqnarray*}

On the other hand, using Theorem \ref{epsilonthm} in Appendix A we have
\begin{eqnarray*}
P(S_{\mathbf{u}}|\text{ }\mathbf{u}) &=&P(n(\sigma ^{2}-\epsilon _{n})\leq
\left\Vert \mathbf{Y}-\mathbf{u}\right\Vert _{2}\}\leq n(\sigma
^{2}+\epsilon _{n})\text{ }|\text{ }\mathbf{u}) \\
&=&1-2\Phi (\frac{-\sqrt{n}\epsilon _{n}}{\sqrt{2}\sigma ^{2}}).
\end{eqnarray*}
Therefore

\begin{eqnarray*}
&& P(D_\mathbf{u}|\mathbf{u})\\
&=&P(S_\mathbf{u}\cap Q_\mathbf{u} |\mathbf{u}) \\
&\geq&P(S_{\mathbf{u}}|\text{ }\mathbf{u})-P(\mathbb{R}^{n}-Q_{\mathbf{u}}|%
\text{ }\mathbf{u}) \\
&=&1-2\Phi (\frac{-\sqrt{n}\epsilon _{n}}{\sqrt{2}\sigma ^{2}})-2\lceil\frac{%
(1/4-b)\log n }{\log k }\rceil\Phi (-\log n).
\end{eqnarray*}

Let $b\rightarrow 0$ and $k\rightarrow \infty$, for $\epsilon
_{n}=n^{-1/2}\log n$, using Appendix A, Theorem \ref{epsilonthm}, clearly we
have

\begin{equation*}
P(D_\mathbf{u}|\mathbf{u})=1-o(1).
\end{equation*}

\textbf{Type II probability error: }

Consider two arbitrary codewords $\mathbf{u}_{1}, \mathbf{u}_{2}$ in $%
\mathcal{U}$. We are going to show that $P(D_{\mathbf{u}_{1}}|\mathbf{u}%
_{2})=o(1)$.

If two codewords $\mathbf{u}_{1}, \mathbf{u}_{2}$ lie in different Galaxies
of depth $\overline{t}=\lceil\frac{(1/4-b)\log n}{\log k }\rceil$ (this
means that $\mathbf{u}_{1}\in A^{\overline{t}}_{\mathbf{o}_{i}}(r,\theta ), 
\mathbf{u}_{2}\in A^{\overline{t}}_{\mathbf{o}_{j}}(r,\theta)$ for some $%
i,j\in \mathcal{M}$ with $i\neq j$), using Lemma \ref{diffgalax}, we have

\begin{equation*}
\|\mathbf{u}_{1}-\mathbf{u}_{2}\|_{2}\geq \frac{n^{b+1/4}}{2}.
\end{equation*}

Therefore by Corollary \ref{epsilcon} Appendix A , we have%
\begin{eqnarray*}
P(S_{\mathbf{u}_{1}}|\text{ }\mathbf{u}_{2}) =o(1)
\end{eqnarray*}

Hence 
\begin{equation*}
P(D_{\mathbf{u}_{1}}|\text{ }\mathbf{u}_{2})\leq P(S_{\mathbf{u}_{1}}|\text{ 
}\mathbf{u}_{2})=o(1)
\end{equation*}
and we are done. Now with no loss of generality assume that $\mathbf{u}_{1}, 
\mathbf{u}_{2}$ are in the Galaxy of depth $\overline{t}$ and center $%
\mathbf{O}_{i}$ for $i\in \mathcal{M}$. For $\theta=2\arcsin(\frac{2}{\sqrt{%
k-2}})$ one can easily see that we have $(\sin(\theta/2)-\frac{1}{k-1})^{2}>%
\frac{1}{k-1}$ and so using Corollary \ref{error2}, $P(P_{\bar{\mathbf{o}},%
\mathbf{u}_{1}}|\mathbf{u}_{2})\leq \lambda _{2}$.

\bigskip

\section{Proof of Theorem~\ref{mainresult}}\label{proof}


\textbf{Proof: }For each $n$ consider an Galaxy code with parameters $%
\mathrm{G}_{n}(\theta,b,k)$ (which is defined in Section 2) where $%
\theta=2\arcsin(\frac{2}{\sqrt{k-2}})$. As we showed in the previous
sections $\mathrm{G}_{n}(\theta,b,k)$ is a Deterministic Identification code
a Gaussian AWGN channel with power constraints. If the number of codewords
in this code is $N_n=2^{nR_n\log n}$, then using Lemma \ref{1} in Section V we have

\begin{equation*}
\frac{\log \sqrt{nP}-1}{\log n}-(-b+1/4)\frac{\log(\sin(\theta))}{\log k }%
-(b+1/4)\leq R_n.
\end{equation*}

Since $\theta=2\arcsin(\frac{2}{\sqrt{k-2}})$, we have $\sin(\theta)=\frac{4%
\sqrt{k-6}}{k-2}<\frac{4}{\sqrt{k}}$ and so for $b<1/4$ we have

\begin{align*}
R_n & \geq & \frac{\log \sqrt{nP}-1}{\log n}-(-b+1/4)\frac{\log(\frac{4}{\sqrt{k}}%
)}{\log(k)}-(b+1/4)\\
& \geq&
\frac{1/2\log n+1/2\log P -1}{\log n}-\\
& &(-b+1/4)\frac{\log 4-1/2\log k }{\log
k }-(b+1/4)\\
& = &\frac{3}{8}+\frac{1/2\log P -1}{\log n}+b(\frac{\log 4}{\log k }-3/2)-\frac{%
\log 4}{4\log k }
\end{align*}
Therefore $R_n$ tends to $\frac{3}{8}+b(\frac{\log 4}{\log k }-3/2)-\frac{%
\log 4}{4\log k }$ when $n$ goes to infinity, which means that we construct $%
DI$ codes where their rates tends to $\frac{3}{8}+b(\frac{\log 4}{\log k }%
-3/2)-\frac{\log 4}{4\log k }$. Since $b$ is a small arbitrary real number
and $k$ is an arbitrary large natural number if $b\rightarrow 0$ and $%
k\rightarrow \infty$, we conclude that $C_{DI}(\mathcal{G})\geq \frac{3}{8}$
and so we are done. 

\section{Conclusions}\label{conclusions}

In this work, we developed a new coding method and demonstrated that it can be used to improve the lower bound on the achievable rate for deterministic identification codes over Gaussian channels with power constraints. 
The core idea of our coding method is as follows: 
All code points are positioned on a hypersphere with a certain minimum angular separation. 
We use a hyperplane, perpendicular to the radius of the point in question, to separate it from the other points. 
This separation is achieved by projecting onto the radius line of the point. 
The method works because, due to the angular separation, the projections of the other points are at a sufficiently large distance from the decision boundary where the hyperplane passes (but on opposite sides). 
Gaussian tail probability bounds handle the remaining analysis. 
When described this way, the scheme scales efficiently through multiple levels.

We were able to prove that the lower bound can indeed be improved to $\frac 38$.
We conjecture that the current upper bound can be achieved. 

\section*{Acknowledgments}\small 
The authors gratefully acknowledge support from the German Federal Ministry of Education and Research (BMBF) through various national initiatives: 

\begin{itemize}
    \item \textbf{6G Communication Systems} via the research hub \textit{6G-life} (Grants 16KISK002 and 16KISK263).
    \item \textbf{Post Shannon Communication (NewCom)} (Grants 16KIS1003K and 16KIS1005).
    \item \textbf{QuaPhySI – Quantum Physical Layer Service Integration} (Grants 16KISQ1598K and 16KIS2234).
    \item \textbf{QTOK – Quantum Tokens for Secure Authentication in Theory and Practice} (Grants 16KISQ038 and 16KISQ039).
    \item \textbf{QUIET – Quantum Internet of Things} (Grants 16KISQ093 and 16KISQ0170).
    \item \textbf{Q.TREX – Resilience for the Quantum Internet} (Grants 16KISR026 and 16KISR038).
    \item \textbf{QDCamnetz – Quantum Wireless Campus Network} (Grants 16KISQ077 and 16KISQ169).
    \item \textbf{QR.X – Quantum link extension} (Grants 16KISQ020 and 16KISQ028).
\end{itemize}

Additionally, the authors acknowledge support from the \textit{6GQT} initiative, funded by the Bavarian State Ministry of Economic Affairs, Regional Development, and Energy.

\section*{Appendix A}

\begin{Theorem}
\label{epsilonthm} If $\mathbf{u}_{i}\in \mathbb{R}^n$ and

\begin{equation*}
S_{\mathbf{u}_{i}}=\{\text{ }\mathbf{y}\in \mathbb{R}^;\text{ }n(\sigma
^{2}-\epsilon _{n})\leq \left\Vert \mathbf{y}-\mathbf{u}_{i}\right\Vert
_{2}\leq n(\sigma ^{2}+\epsilon _{n})\text{ }\}.
\end{equation*}

Then for large enough $n$, we have

\begin{eqnarray*}
P(S_{\mathbf{u}_{i}}|~\mathbf{u}_{i})&=&1-2\Phi (\frac{-\sqrt{n}\epsilon _{n}%
}{\sqrt{2}\sigma ^{2}}) \\
&=&1-o(e^{-n\epsilon_{n}^{2}/4\sigma ^{4}}).
\end{eqnarray*}
\end{Theorem}

\bigskip

\textbf{Proof:} Note that if the codeword $\mathbf{u}_{i}$ has been sent,
then for $k=1,\ldots ,n$, the variables $Z_{k}=\frac{Y_{k}-\mathbf{u}_{ik}}{%
\sigma },$ are independent normal random variables $\mathcal{N}(0,1)$ and
then $\sum\limits_{k=1}^{n}Z_{k}^{2}$ has Chi-Square distribution with $n$
degree of freedom, denoted by $\chi ^{2}(n)$. Therefore,

\bigskip

\begin{eqnarray*}
&& P(S_{\mathbf{u}_{i}}|\text{ }\mathbf{u}_{i})\\
&=&P(n(\sigma ^{2}-\epsilon
_{n})\leq \left\Vert \mathbf{Y}-\mathbf{u}_{i}\right\Vert _{2}\}\leq
n(\sigma ^{2}+\epsilon _{n})\text{ }|\text{ }\mathbf{u}_{i}) \\
&=&P(n-\frac{n\epsilon _{n}}{\sigma ^{2}}\leq
\sum\limits_{k=1}^{n}Z_{k}^{2}\leq n+\frac{n\epsilon _{n}}{\sigma ^{2}}) \\
&=&P(-\frac{\sqrt{n}\epsilon _{n}}{\sqrt{2}\sigma ^{2}}\leq \frac{\chi
^{2}(n)-n}{\sqrt{2n}}\leq +\frac{\sqrt{n}\epsilon _{n}}{\sqrt{2}\sigma ^{2}})
\\
&\overset{a}{\simeq }&\Phi (\frac{\sqrt{n}\epsilon _{n}}{\sqrt{2}\sigma ^{2}}%
)-\Phi (\frac{-\sqrt{n}\epsilon _{n}}{\sqrt{2}\sigma ^{2}}) \\
&=&1-2\Phi (\frac{-\sqrt{n}\epsilon _{n}}{\sqrt{2}\sigma ^{2}})
\end{eqnarray*}

\bigskip Approximation (a) is true because if $n$ is large enough, the
random variable $\frac{\chi ^{2}(n)-n}{\sqrt{2n}}$ has standard normal
distribution approximately.

Besides, by Mill's inequality,

\bigskip 
\begin{equation*}
\Phi (\frac{-\sqrt{n}\epsilon _{n}}{\sqrt{2}\sigma ^{2}})\leq \frac{2\sigma
^{2}}{\sqrt{n\pi \text{ \ }}\epsilon _{n}}e^{-(\sqrt{n}\epsilon
_{n})^{2}/4\sigma ^{4}},
\end{equation*}

which goes to zero if $\sqrt{n}\epsilon _{n}\longrightarrow \infty$. Then we
have

\begin{equation*}
P(S_{\mathbf{u}_{i}}|\text{ }\mathbf{u}_{i})\geq 1-\frac{2\sigma ^{2}}{\sqrt{%
n\pi \text{ \ }}\epsilon _{n}}e^{-(\sqrt{n}\epsilon _{n})^{2}/4\sigma ^{4}}.
\end{equation*}%
Therefore we have $P(S_{\mathbf{u}_{i}}|\text{ }\mathbf{u}%
_{i})=1-o(e^{-n\epsilon _{n}^{2}/4\sigma ^{4}}\text{ })$, which proves the
theorem.

\bigskip

\begin{Remark}
Therefore $\ P(S_{\mathbf{u}_{i}}|$ $\mathbf{u}_{i})\geq 1-\lambda _{1}$ is
equivalent \ with \ $\frac{\sqrt{n}\epsilon _{n}}{\sqrt{2}\sigma ^{2}}\geq
\Phi ^{-1}(1-\lambda _{1}/2).$ In other words we need to choose
\end{Remark}

\begin{equation}
\sqrt{n}\epsilon _{n}\geq \sqrt{2}\sigma ^{2}\Phi ^{-1}(1-\lambda _{1}/2). 
\tag{1}
\end{equation}%
and when $\lambda _{1}$ goes to zero, $\Phi ^{-1}(1-\lambda _{1}/2)$ goes to
infinity very fast.

\begin{Theorem}
Consider two codewords $\mathbf{u}_{i}$ and $\mathbf{u}_{j},$ with distance $%
d_{n}=\left\Vert \mathbf{u}_{i}-\mathbf{u}_{j}\right\Vert _{2}$. Then for
large enough $n$, we have 
\begin{equation*}
P(S_{\mathbf{u}_{j}}|\text{ }\mathbf{u}_{i})\simeq \Phi (\frac{n\epsilon
_{n}-d_{n}^{2}}{\sigma \sqrt{2n\sigma ^{2}+4d_{n}{}^{2}}})
\end{equation*}
\end{Theorem}

\bigskip \bigskip \textbf{Proof:} \bigskip For the first step, we assume $%
\mathbf{u}_{i}=\mathbf{o}=(0,0,\cdots ,0)$ and $\mathbf{u}%
_{j}=(d_{n},0,\cdots ,0)$. Therefore we have
\begin{eqnarray*}
&& P(S_{\mathbf{u}_{j}}|\text{ }\mathbf{u}_{i})\\ 
&=&P(n\sigma ^{2}-n\epsilon
_{n}\leq \left\Vert \mathbf{Y}-\mathbf{u}_{j}\right\Vert _{2}\}\leq n\sigma
^{2}+n\epsilon _{n}\text{ }|\text{ }\mathbf{u}_{i}) \\
&=&P(n\sigma ^{2}-n\epsilon _{n}\leq (\sigma Z_{1}-d_{n})^{2}+\sigma
^{2}\sum\limits_{k=2}^{n}Z_{k}^{2}\leq n\sigma ^{2}+n\epsilon _{n}) \\
&=&P(n-\frac{n\epsilon _{n}}{\sigma ^{2}}-\frac{d_{n}^{2}}{\sigma ^{2}}\leq
\sum\limits_{k=1}^{n}Z_{k}^{2}-2\frac{d_{n}}{\sigma }Z_{1}\leq n+\frac{%
n\epsilon _{n}}{\sigma ^{2}}-\frac{d_{n}^{2}}{\sigma ^{2}}).
\end{eqnarray*}%
On the other hand, $\sum\limits_{k=1}^{n}Z_{k}^{2}\thicksim {\large \chi }%
^{2}(n),$ and for large $n$, $\sum\limits_{k=1}^{n}Z_{k}^{2}\thickapprox 
\mathcal{N}(n,2n)$. Also, $-2\frac{d_{n}}{\sigma }Z_{1}\thicksim \mathcal{N}%
(0,4d_{n}{}^{2}/\sigma ^{2})$ and $Cov(\sum\limits_{k=1}^{n}Z_{k}^{2},-2%
\frac{d_{n}}{\sigma }Z_{1})=0$, which implies $\sum%
\limits_{k=1}^{n}Z_{k}^{2}-2\frac{d_{n}}{\sigma }Z_{1}\thickapprox \mathcal{N%
}(n,2n+4d_{n}{}^{2}/\sigma ^{2}),$ and then%
\begin{eqnarray*}
P(S_{\mathbf{u}_{j}}|\text{ }\mathbf{u}_{i}) &=&\Phi (\frac{n\epsilon
_{n}-d_{n}^{2}}{\sigma \sqrt{2n\sigma ^{2}+4d_{n}{}^{2}}}{})-\Phi (\frac{%
-n\epsilon _{n}-d_{n}^{2}}{\sigma \sqrt{2n\sigma ^{2}+4d_{n}{}^{2}}}) \\
&\thickapprox &\Phi (\frac{n\epsilon _{n}-d_{n}^{2}}{\sigma \sqrt{2n\sigma
^{2}+4d_{n}{}^{2}}})
\end{eqnarray*}%

Step 2. Let $\mathbf{u}_{i}=\mathbf{o}=(0,0,...,0)$ and $\mathbf{u}%
_{j}=(a_{1},a_{2},\cdots ,a_{n}),$ be two codewords with distance $d_{n}=%
\sqrt{\sum\limits_{k=1}^{n}a_{k}^{2}}=\left\Vert \mathbf{u}_{i}-\mathbf{u}%
_{j}\right\Vert _{2}\geq n^{\frac{1}{4}}\log n $.%
\begin{eqnarray*}
&& P(S_{\mathbf{u}_{j}}|\mathbf{u}_{i}) \\
&=&P(n\sigma ^{2}-n\epsilon _{n}\leq
\left\Vert \mathbf{Y}-\mathbf{u}_{j}\right\Vert _{2}\}\leq n\sigma
^{2}+n\epsilon _{n}\text{ }|\text{ }\mathbf{u}_{i}) \\
&=&P(n\sigma ^{2}-n\epsilon _{n}\leq \sigma
^{2}\sum\limits_{k=1}^{n}(Z_{k}-a_{k}/\sigma )^{2}\leq n\sigma
^{2}+n\epsilon _{n}) \\
&=&P(n-n\epsilon _{n}/\sigma ^{2}\leq \sum\limits_{k=1}^{n}Z_{k}^{2}-\frac{2%
}{\sigma }\sum\limits_{k=1}^{n}a_{k}Z_{k}+d_{n}{}^{2}/\sigma ^{2}\\
&&\leq
n+n\epsilon _{n}/\sigma ^{2}) \\
&=&P(n-n\epsilon _{n}/\sigma ^{2}-d_{n}{}^{2}/\sigma ^{2}\leq
\sum\limits_{k=1}^{n}Z_{k}^{2}-\frac{2}{\sigma }\sum%
\limits_{k=1}^{n}a_{k}Z_{k}\\
&&\leq n+n\epsilon _{n}/\sigma
^{2}-d_{n}{}^{2}/\sigma ^{2})
\end{eqnarray*}

Since we have $Cov(\sum\limits_{k=1}^{n}Z_{k}^{2},\sum%
\limits_{k=1}^{n}a_{k}Z_{k})=0$, the same result will be obtained.

\bigskip Step 3. For two arbitrary codewords $\mathbf{u}_{i}$ and $\mathbf{u}%
_{j}$, we can use the previous steps by the following property to complete
our proof.%
\begin{equation*}
P(S_{\mathbf{u}_{j}}|\mathbf{u}_{i})=P(S_{\mathbf{u}_{j}-\mathbf{u}_{i}}|%
\mathbf{o}).
\end{equation*}

Therefore we have the following corollary.

\begin{Corollary}
\label{epsilcon} Then for $\epsilon _{n}=\frac{\log (n)}{\sqrt{n}},d_{n}\geq
n^{\frac{1}{4}}\log (n)$ and large enough $n$, we have 
\begin{equation*}
P(S_{\mathbf{u}_{j}}|\text{ }\mathbf{u}_{i})=o(1)
\end{equation*}
\end{Corollary}

\bigskip

\section*{Appendix B}

The inner product of two arbitrary vectors $\vec{\mathbf{v}}$ and $\vec{%
\mathbf{u}}$ in $\mathbb{R}^{n}$, is defined by $<\vec{\mathbf{u}},\vec{%
\mathbf{v}}>=\sum\limits_{i=1}^{n}u_{i}v_{i}$. Clearly $\vec{\mathbf{v}}%
-Proj_{\vec{\mathbf{u}}}(\vec{\mathbf{v}})$ and $\vec{\mathbf{u}}$ are
orthogonal, $(\vec{\mathbf{v}}-Proj_{\vec{\mathbf{u}}}(\vec{\mathbf{v}}%
))\perp \vec{\mathbf{u}}$ (which means that their inner product is zero) and
then 
\begin{equation*}
<\vec{\mathbf{u}},\vec{\mathbf{v}}>=<\vec{\mathbf{u}},Proj_{\vec{\mathbf{u}}%
}(\vec{\mathbf{v}})>
\end{equation*}%
and%
\begin{equation*}
\left\vert <\vec{\mathbf{u}},Proj_{\vec{\mathbf{u}}}(\vec{\mathbf{v}}%
)>\right\vert =\left\Vert \mathbf{u}\right\Vert_2 \left\Vert Proj_{\vec{%
\mathbf{u}}}(\vec{\mathbf{v}})\right\Vert_2.
\end{equation*}
Therefore%
\begin{equation*}
\left\Vert Proj_{\vec{\mathbf{u}}}(\vec{\mathbf{v}})\right\Vert_2 =\frac{%
\left\vert <\vec{\mathbf{u}},\vec{\mathbf{v}}>\right\vert }{\left\Vert \vec{%
\mathbf{u}}\right\Vert_2 }.
\end{equation*}

For $\vec{\mathbf{Z}}=(Z_{1},Z_{2},...,Z_{n})$, where $Z_{1},Z_{2},...,Z_{n}$
are iid random variables with standard normal distribution, we have
\begin{equation*}
<\vec{\mathbf{u}},\vec{\mathbf{Z}}>=\sum\limits_{i=1}^{n}u_{i}Z_{i}\thicksim
N(0,\left\Vert \vec{\mathbf{u}}\right\Vert_{2} ^{2}).
\end{equation*}%
Therefore%
\begin{eqnarray*}
P(\left\Vert Proj_{\vec{\mathbf{u}}}(\vec{\mathbf{Z}})\right\Vert_2 &\leq& x
)=P(\frac{\left\vert <\vec{\mathbf{u}},\vec{\mathbf{Z}}>\right\vert }{%
\left\Vert \vec{\mathbf{u}}\right\Vert_2 }\leq x) \\
&=&\Phi (x)-\Phi (-x) \\
&=&1-2\Phi (-x),
\end{eqnarray*}

equivalently we have

\begin{eqnarray*}
P(\left\Vert Proj_{\vec{\mathbf{u}}}(\vec{\mathbf{Z}})\right\Vert_2 &\geq& x
)=2\Phi (-x).
\end{eqnarray*}

\bibliographystyle{IEEEtran}
\bibliography{references}

\end{document}